# INFORMATION HIDING IN CSS: A SECURE SCHEME TEXT-STEGANOGRAPHY USING PUBLIC KEY CRYPTOSYSTEM

Herman Kabetta[1], B. Yudi Dwiandiyanta[2], Suyoto[3]

Department of Informatics Engineering, Atma Jaya Yogyakarta University, Yogyakarta, Indonesia

[1] betabox@rocketmail.com
[2] yudi-dwi@staff.uajy.ac.id
[3] suyoto@staff.yajy.ac.id

## ABSTRACT

*In many recent years, the programming world has been introduced about a new programming language for designing websites, it is CSS that can be be used together with HTML to develop a web interface. And now, these two programming languages as if inseparably from each other. As a client-side scripting, CSS is visible by all users as the original script, but it can not be granted changed. Website is a tool of information disseminator throughout the world, this is certainly can be used to a secret communication by using CSS as a message hider. This paper proposed a new scheme using web tools like CSS for hiding informations. This is a secret communication mechanism using text steganography techniques that is embedded messages on CSS files and is further encrypted using RSA as a public key cryptographic algorithm.*

## 1. INTRODUCTION

Secret communication scheme has been known long time ago, Julius Caesar used cryptography to encode his political directions [1]. Steganography (usually called as stego), the art of hiding message, has been used for many generations. Steganography is often difficult to distinguish with cryptographic because of similarities these functions areas in terms of protecting critical information. The difference between these two method is in terms of how to protect informations. Steganography to disguising the information on the other media so that people do not feel the existence of such informations behind [2]. Meanwhile cryptography protects data by altering information into a form that unreadable or cannot understood by unauthorized people [3]. But, sometimes steganography used in combination with cryptography that offer privacy and security are higher through the communication channel [4].

### 1.2. Information Hiding

The main purpose of steganography is to hiding information in the media which covered so that outsiders will not find any information contained behind [5], figure 1 shows the model of information hiding. Most implementations of steganography performed on the image [4][6] and sound [7][8]. A simple approach to embedding information on the image and sound is by inserting a message into the low bit (Least Significant Bits / LSB) in the pixel data which make up 24 bit BMP image file or between frames (BF) in the MP3 file [9]. Text steganography is the most difficult kind of steganography [5]; this is due largely to the relative lack of redundant information in a text file as compared with a picture or a sound file [10].

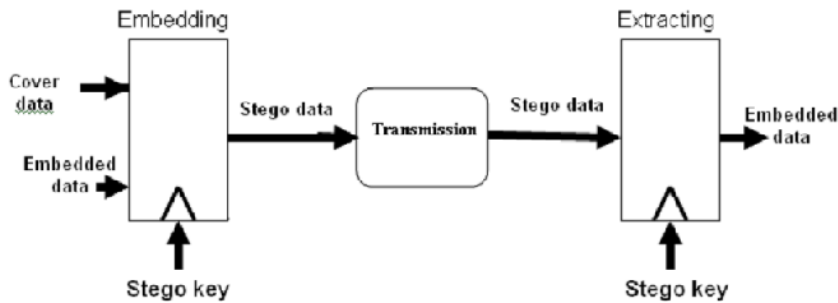

Figure 1. Model for Information Hiding [22]

Steganography is derived from Greek, Steganos (Covered) and Graptos (Writing) [11][12]. It's technically means that the message that is covered or hidden. History records about the steganography implementation was in Ancient Greece, they used to select messengers and shave their head, and then write a message on their head. Once the message had been written the hair was allowed to grow back.After the hair grew back the messenger was sent to deliver the message, the recipient would shave off the messengers hair to see the secret message.

Another method used in Greece was where someone would peel wax off a tablet that was covered in wax, write a message underneath the wax then re-apply the wax. The recipient of the message would simply remove the wax from the tablet to view the message. Even during World War I and II, invisible ink used for write information on paper sheet so that only the pieces of blank paper that is not suspicious [13]. Until the modern era, as now, even terrorist use steganography for communication, several government sources even suspect that Osama bin Laden videotape that plays on television stations throughout the world was contain the hidden messages [14] .

According to Nosrati et al (2011), that are three categories of file formats that can be used for steganography techniques, text, image and audio as shown in figure 2.

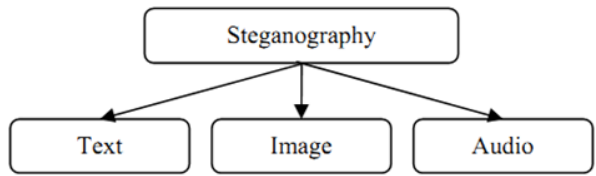

Figure 2. Steganography Types Diagram [15]

## 1.2. Text Steganography

Text steganography is broadly classified into the two categories; Linguistic Steganography, which is further divided into semantic and syntactic method. Another category is Format based steganography which is further divided into following categories, line-shift encoding, word-shift encoding, open-space encoding and feature encoding as described in the figure 3.

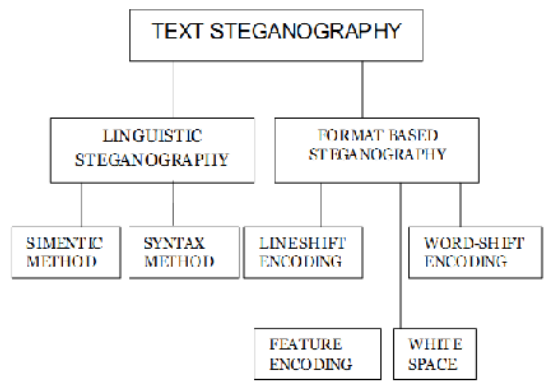

Figure 3. Types of Text Steganography [1]

### 1.3. RSA as An Assymmetric Key Cryptosystem

There are two kinds of cryptographic algorithms based on the key that is used for encryption and decryption [16], there are symmetric algorithm and asymmetric algorithms. In symmetric algorithm, key for encryption is same with a key for decryption, because that is called symmetric cryptography. Whereas in asssymmetric algorithm, there are different key for encryption and decryption, public key is for encryption and private key for decryption. The first inventor of asymmetric key cryptography algorithm is Clifford Cocks, James H. Ellis and Malcolm Williamson (a group of mathematicians who worked for United Kindom's Government Communications Head Quarters, the British secret agent) at the beginning of the year 1970 [17].

RSA is a popular one than another assymmetric-key cryptographic systems [18]. The security of the RSA cryptosystem relies on the believed difficulty of factoring large composite integers [19]. The RSA algorithm is named after Ron Rivest, Adi Shamir and Len Adleman. It consists of the following procedures: key generation, encryption, decryption [20].

#### 1.3.1. Key Generation

1. Choose two big primes: p and q.
2. Calculate n=p*q.
3. Randomly choose an integer e, satisfying $1<e< \phi(n)$, gcd (e, $\phi(n)$) = 1. Totient function $\phi(n)$ denotes the number of positive integers less than n and relatively prime to n. Here $\phi(n) = (p-1)*(q-1)$. The public key is (e, n).
4. Calculate d, satisfying $ed \bmod \phi(n) = 1$ , the private key is (d, n).

#### 1.3.2. Encryption Procedure

1. Partition the message m to groups $m_i$, i=1,2,…, | $m_i$ |= |n|-1; (|a| means the length of a in binary form) .
2. Encrypt each group: $c_i = m_i^e \bmod n$.
3. Connect each $c_i$ and get the cipher text c.

#### 1.3.3. Decryption Procedure

1. Partition c to $c_i$, i=1,2,…, | $c_i$ |= |n|-1;
2. Decrypt each $c_i$ : $m_i = c_i^d \bmod n$
3. Connect each $m_i$ and recover the plain text m.

### 1.4. Cascading Style Sheet (CSS)

Cascading Style Sheets (CSS) is a style sheet language used to describe the presentation semantics (the look and formatting) of a document written in a markup language. Its most common application is to style web pages written in HTML and XHTML. CSS has a simple syntax and uses a number of English keywords to specify the names of various style properties. A style sheet consists of a list of rules. Each rule or rule-set consists of one or more selectors and a declaration block. A declaration-block consists of a list of declarations in braces. Each declaration itself consists of a property, a colon (:), a value. If there are multiple declarations in a block, a semi-colon (;) must be inserted to separate each declaration.

Here is an example summing up the rules above :

Selector [, Selector2, ...] [:Pseudo-class] {
Property: Value;
}

Below is the example of CSS based on the rules above :

h1 {
color: white;
background-color: orange !important;
}

Several programmers written CSS syntax in one line for each selector, and therefore usage the simply "End of Line" technique will make the CSS looks more cluttered and eventually could be suspected by any third parties.

### 1.5. An Overview of A Proposed Scheme

The idea of embedding a secret message into a CSS file (Cascading Style Sheet) is inspired by research conducted by Por and Delina [21] and Memon et al [22]. Hiding a message was performed into XML data [22], now I assumed it can be applied to the CSS file by embedding the message into the each CSS style properties too, ie after semi-colon characters. More precisely using the "End of Line Spacing" that is using whitespaces to encode the message by inserting spaces or tabs in the end of the line of CSS style properties.

A new approach for sending messages without any fear that the message is intercepted and then modified by an intruder that will sends a false message to the recipient. Figure 4 shows the model of proposed scheme, by using a public key cryptosystem can also provide more secure that messages is more difficult to be solved. It can be done because of steganography that combined with public key cryptosystem has also been carried out by Bandyopadhyay into his research [23] , but its using images as a cover media.

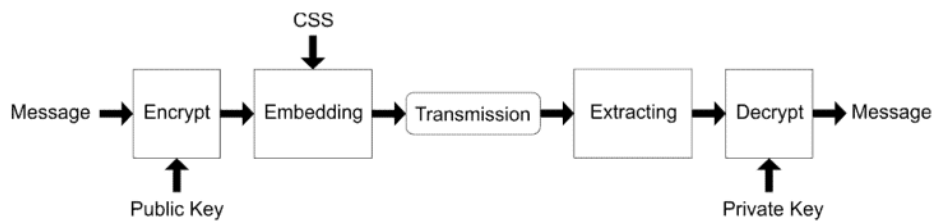

Figure 4. Model for A Proposed Scheme

## 2. Previous Works

Research about Steganography has been widely performed, especially steganography into the image data. Bandyopadhyay and Chakraborty conducted their research in 2011 [24], their study elaborated about an embedding message into the images using the four algorithms which adjusted in the DNA sequence. In the previous year Bandyopadhyay et al (2010) [25] also has introduced a new steganographic method based on genetic algorithm on his research. While combined of steganography and cryptography has also been carried out simultaneously, such as in 2010, Narayana and Prasad [26] introduces a new approach to securely embedding into the image steganography using cryptographic techniques and conversion, this method shows how to secure the image by converting it into image cipher using a secret key algorithm S-DES and hiding image in another image using steganography methods. The proposed method also prevents the possibility of steganalysis.

Steganography on other media such as digital sound has also been conducted, Geetha and Muthu in their research (2010) about the audio steganography said that embedding a secret message into digital sound is very difficult to perform than the digital image [27], this is due to not presence of additional bytes generated for embedding[28]. In audio steganography, a secret message embedded by slightly altering the binary sequence of sound files. In a research that conducted by Atoum et al (2011) [7][8], proposes a new method of embedding information into the audio media files (MP3) between frames (BF) in MP3 files.

Most research of steganography was using cover media such as images, video clips and sound. However, steganography into the text is usually not preferred because of the difficulty in finding redundant bits in a text document [1]. To embedding an information into a text document, its characteristics must be changed first. This characteristic can be either text format or characteristics of the characters. But the problem is that when a small change have been made to the texy document it will be visible by intruders or attackers. Some of the methods proposed to solve the problem, such as by line shifting, words shifting, up to whitespaces manipulation into the cover text [29].

Banerjee et al (2011) conducted research on text steganography, on his paper [30], Banerjee introduces a method of hiding messages into the text by changing the prefix "a" or "an" into the cover of the English text. A new approach is proposed in hiding information using inter-words spacing and line spacing between paragraphs as a hybrid method, Por et al (2008) called it "Whitesteg" [4].

The schematic used on Whitesteg is by converting the secret message into a binary bits of each message which then embedded into any whitespace in the text cover by changing it first into the another whitespaces characters, single-spaced for "0" and two spaces for "1 " [4][21]. Disadvantages of this method is the limited amount of message bits to be embedded depending on the number of bits of text that is used as a cover, and has not used the encryption systems to secure the message that embedded, so the message will be easily solved when stego text has been suspected containing a secret message by intruders.

A study providing a new horizon for securely communications through XML on the Internet [22]. XML allows the sending the message can not be changed when it is intercepted by an intruder. The scheme that is used is embedding the message between the XML tags. But the shortcomings of this scheme is has not used a cryptographic techniques to encrypt the message so that if the intruder knows the techniques of the steganographic, then the message can be solved. Different from the papers by Mir and Hussain (2010), which discusses steganography text through XML with a secret message that is encrypted using cryptographic algorithm AES (Advanced Encryption Standard) first [31].

## 3. CSS STEGANOGRAPHY SCHEME

Web based communication has a great amount of bandwidth and hence can be used for secret communication. HTML and CSS are two basic but important and universal tools for web development. This paper proposes a new scheme on hiding information that is embedded through a Cascading Style Sheet (CSS) by using End Of Line (EOL) on each CSS style properties, exactly after a semi-colon. Before embedded into the cover text, message firstly encrypted using RSA Algorithm, and then it transmitted to the receiver.

In the literature [4], hiding information within spaces appears to have potential as people can hardly identify the existence of the hidden bits which appear in the whitespaces between the words. Por et al. had shown that one space is interpreted as "0" whereas two spaces are interpreted as "1". But using two spaces between the words, it would make stegotext more suspicious. Overcome for this problem hence one the best ways is by hiding information at the end of the line. Mir and Hussain had shown that it was applied on XML files and was further encrypted using Advanced Ecryption Standard (AES) [31]. But it is still not secure, as we know, the problems of symmetric-key algorithm is about the key distribution [32]. Therefore, this proposed scheme, as shown in figure 5, would be applied an asymmetric-key algorithm for the encryption process.

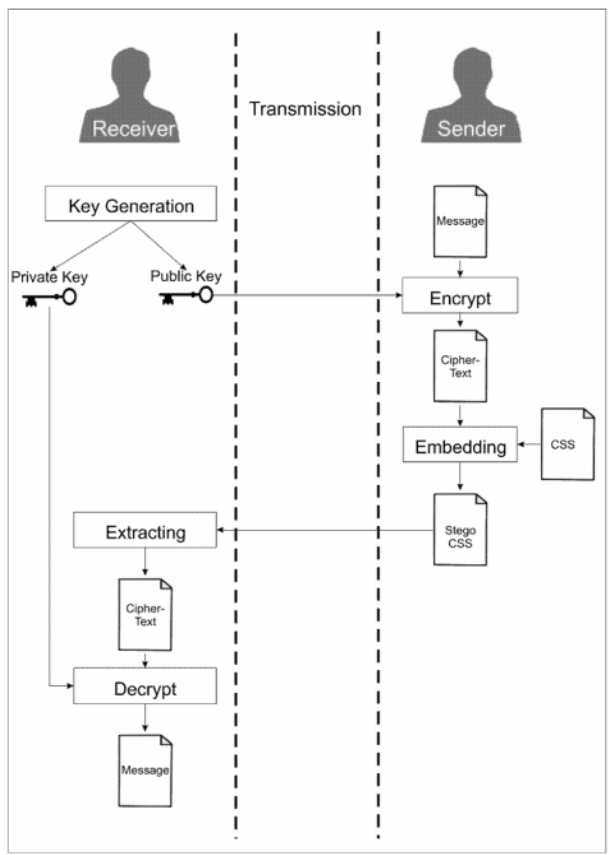

Figure 5. System Flow of A Proposed Scheme

As follow the explanation of figure 5:
Step 1, Receiver generate with RSA Algorithm for a pair of key, that is private and public key, as shown in figure 6 (left).
Step 2, Receiver sends public key to the Sender.

Step 3, Sender encrypt the secret message with RSA Algorithm, then the ciphertext embedding into CSS file. The result of this step is a stego CSS as shown in figure 7.

Step 3a, encrypted message with RSA Algorithm.

Step 3b, converting ciphertext to biner.

Step 3c, converting biner to whitespace, space for 0 and tab for 1.

Step 3d, searching for semi-colon then inserting the whitespaces in the after character of semi-colon.

Step 4, Sender sending the stego CSS to the Receiver.

Step 5, Receiver get the stego CSS and extract it for getting the cipher text, then Deciphering the cipher text to find out the secret message, this step shown in figure 6 (Right).

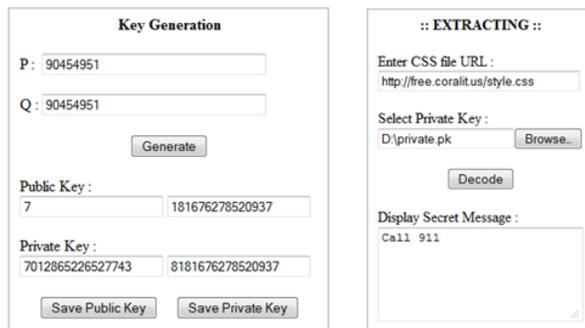

Figure 6. GUI for Receiver: Key Generation (Left), Extracting Process (Right)

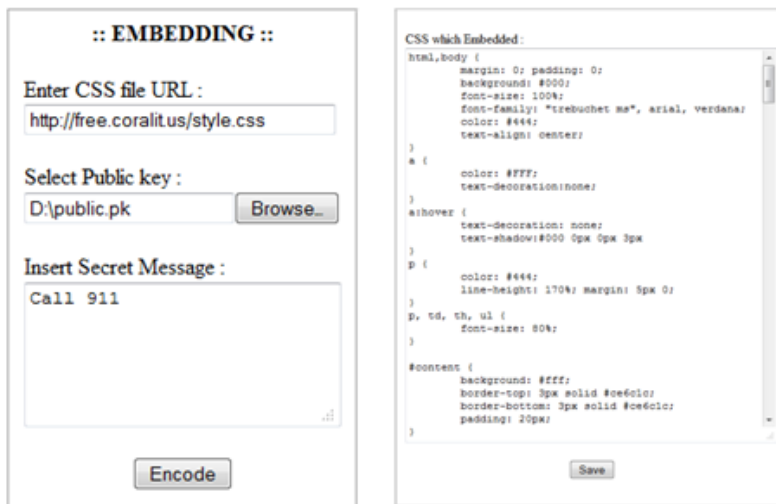

Figure 7. GUI for Sender : Embedding Process

## 3. CONCLUSIONS

The new approach uses this scheme shows that the stego text looks as same as the original text, by using the "End of Line" techniques for the embedding process makes no obvious changes as shown in figure 8. Usage of the public key cryptography is also increasing the security of hidden information. Since CSS stored on the webserver so it is not possible to changing the data by any third parties. Weakness of this technique is the limited amount of characters that can be embedded, it is depending of the available semi-colon amount.

```
html,body {                                    html,body {
        margin: 0; padding: 0;                         margin: 0; padding: 0;
        background: #000;                              background: #000;
        font-size: 100%;                               font-size: 100%;
        font-family: "trebuchet ms", arial, verdana;   font-family: "trebuchet ms", arial, verdana;
        color: #444;                                   color: #444;
        text-align: center;                            text-align: center;
}                                              }
a {                                            a {

        color: #FFF;                                   color: #FFF;
        text-decoration:none;                          text-decoration:none;

}                                              }
a:hover {                                      a:hover {
        text-decoration: none;                         text-decoration: none;
        text-shadow:#000 0px 0px 3px;                  text-shadow:#000 0px 0px 3px;
}                                              }
p {                                            p {

        color: #444;                                   color: #444;
        line-height: 170%;                             line-height: 170%;
        margin: 5px 0;                                 margin: 5px 0;
}                                              }
p, td, th, ul {                                p, td, th, ul {
        font-size: 80%;                                font-size: 80%;
                                               }
```

Figure 8. Original CSS File (Left), Stego CSS With Hidden Information (Right)

The future work should focus towards the range of the payload size can be increased so that more data is able to be embedded in the CSS files and not only text message can be embedded, but also the image and sound.

**Herman Kabetta**

Herman Kabetta received his Bachelor Degree in Mathematics from the University of General Soedirman (Purwokerto, Indonesia). Now He was studying in Master Degree of Informatics Engineering in Atma Jaya Yogyakarta University.

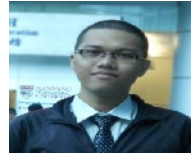

**Yudi Dwiandiyanta**

Yudi Dwiandiyanta, S.T., M.T. is a lecturer in Atma Jaya University, He received his Master Degree in Electrical Engineering from the Gadjah Mada University (Yogyakarta, Indonesia). Several His researches performed on the subject of soft computing and another research that is sponsored by Indonesian government.

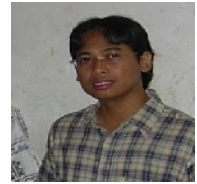

**Suyoto**

Prof. Ir. Suyoto, M.Sc.,Ph.D. is a lecturer in Atma Jaya University, He received his last degree in Universiti Kebangsaan Malaysia. Reviewer of many National and International Conference, one of which is "The 12th International Conference on Information Integration and Web-based Applications & Services (iiWAS2010)".

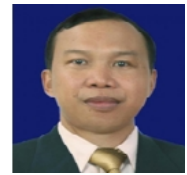